\documentclass[aps,prl,groupedaddress,superscriptaddress,amsmath,amssymbol,stfloats,twocolumn]{revtex4-1}
\usepackage[english]{babel} 
\usepackage{amsfonts}
\usepackage{amsmath}
\usepackage{fullpage}%
\usepackage{amsmath}
\usepackage{bbm}
\usepackage[mathscr]{eucal}
\usepackage{amssymb}%
\usepackage{graphicx}
\providecommand{\U}[1]{\protect\rule{.1in}{.1in}}

\newcommand{\tr}{\mathrm{tr}}

\newtheorem{theorem}{Theorem}

\newtheorem{lemma}[theorem]{Lemma}

\newenvironment{proof}[1][Proof]{\noindent\textbf{#1.} }{\ \rule{0.5em}{0.5em}}
\usepackage{color}

\begin{document}

\title{\textbf{Discrimination power of a quantum detector.}}

\author{Christoph Hirche}
\affiliation{F\'{\i}sica Te\`{o}rica: Informaci\'{o} i Fen\`{o}mens Qu\`{a}ntics, Departament de F\'{i}sica, Universitat Aut\`{o}noma de Barcelona, ES-08193 Bellaterra (Barcelona), Spain}
\author{Masahito Hayashi}
\affiliation{Graduate School of Mathematics, Nagoya University, Nagoya, Japan}
\affiliation{Centre for Quantum Technologies, National University of Singapore, Singapore}
\author{Emilio Bagan}
\affiliation{F\'{\i}sica Te\`{o}rica: Informaci\'{o} i Fen\`{o}mens Qu\`{a}ntics, Departament de F\'{i}sica, Universitat Aut\`{o}noma de Barcelona, ES-08193 Bellaterra (Barcelona), Spain}
\author{John Calsamiglia}
\affiliation{F\'{\i}sica Te\`{o}rica: Informaci\'{o} i Fen\`{o}mens Qu\`{a}ntics, Departament de F\'{i}sica, Universitat Aut\`{o}noma de Barcelona, ES-08193 Bellaterra (Barcelona), Spain}

\begin{abstract}

We investigate the ability of a quantum measurement device to discriminate two states or, generically, two hypothesis. In full generality, the measurement can be performed a number $n$ of times, and arbitrary pre-processing of the states and post-processing of the obtained data is allowed. Even if the two hypothesis correspond to orthogonal states, perfect discrimination is not always possible. There is thus an intrinsic error associated to the measurement device, which we aim to quantify, that limits its discrimination power. We minimize various error probabilities (averaged or constrained)  over all  pairs of $n$-partite input states.
These probabilities, or their exponential rates of decrease in the case of large~$n$,  give measures of the discrimination power of the device.
For the asymptotic rate of the averaged error probability, 
we obtain a Chernoff-type bound, dual to the standard Chernoff bound for which the state pair is fixed and the optimization is over all measurements. The key point in the derivation is that i.i.d. states become optimal in asymptotic settings.  Minimum asymptotic rates are also obtained for constrained error probabilities, dual to Stein's Lemma and Hoeffding's bound.  We further show that adaptive protocols where the state preparer gets feedback from the measurer do not improve the asymptotic rates. These rates thus quantify the ultimate discrimination power of a measurement device.

%
\end{abstract}

\maketitle

Quantum-enabled technologies exploit the laws 
that govern the microscopic world to outperform their classical counterparts. Detectors, or  measurement devices, are a key ingredient in quantum protocols. They are the interface that connects the microscopic world of quantum phenomena and the world of classical, macroscopically distinct, events~that we observe. It is only through measurements that~we~can access the information residing in quantum systems and 
ultimately make use of any quantum advantage. 

We often encounter experimental situations where 
measurement devices (e.g., Stern Gerlach apparatus, heterodyne detectors, photon counters, fluorescence spectrometers)  are a given. A natural question is then to ask about the ability or power of those devices to perform certain quantum information-processing tasks. 
The informational power of a measurement has been addressed in several ways~\cite{OCMB11}, e.g., via the ``intrinsic data" it provides~\cite{w03} or the capacity of the quantum-classical channel it defines~\cite{OCMB11,DDS11,D15,h12,BRW14}, or via some associated entropic quantities~\cite{s14,s15,s16}. 
%

In this letter we focus on what is arguably the most fundamental primitive in quantum information processing: state discrimination, or generically, quantum hypothesis testing. Our aim is to explore how well a quantum measurement device can discriminate two hypotheses. This problem is dual to that of exploring how well two given quantum states can be discriminated~\cite{H1976}. This is of practical interest since preparing states is often easier than tailoring optimal measurements for a given state~pair.


In a generic discrimination protocol the measurement is performed  
not just once but a number~$n$ of times. The measurement device is 
the only means of extracting classical data from the quantum system used in the protocol, but to improve its performance, one is free to apply any trace preserving quantum operation to the system prior to measuring. Likewise, we view data processing also as a free operation. It is then meaningful to ask what is the minimum 
error probability of discrimination over all possible pre- and post-procesings. 

We can go a step further and minimize the error probability over all state pairs. This characterizes an intrinsic limitation on the discrimination performance of the measurement device since, in general, a device cannot perfectly discriminate two hypothesis, not even when they are given by orthogonal states.  This characterization is  of practical relevance since it sets the ultimate limit on the successful identification of two arbitrary states, for instance two critical phases of a quantum many-body system, when one is bound to a given type of measurements apparatus.


Special attention will be paid to asymptotically large $n$.
We will prove that in this regime 
pairs of entangled states provide no advantage over those of i.i.d. states, of the form $\rho^{\otimes n}$. This is in sheer contrast with the dual problem where the measurement is optimized for fixed i.i.d. states; there is strong numerical evidence  that  collective non-separable measurements are required to attain the corresponding optimal exponential rate of the error probability, given by the quantum Chernoff bound~\cite{ACMBMAV07,CMMAB08}. In the proof, we approach state discrimination as a communication problem and allow for adaptive protocols. In this extended hypothesis testing scenario, we show that adaptive protocols perform better than any fixed protocol, including those that use entangled inputs.  A~few examples for small $n$ will be briefly discussed.
%


Before addressing the problem in detail, it will be helpful to 
recall a few definitions and  results concerning hypothesis testing. 
Here, the so called null, $H_0$, and alternative, $H_1$, hypotheses refer respectively to two possible states,~$\rho$, $\sigma$, of a quantum system~${S}$. In~quantum hypothesis testing one is confronted with the task of deciding which  hypothesis holds by performing a measurement on 
$S$. With full generality, the measurement is defined by a two-outcome Positive Operator Valued Measure (POVM), ${\mathscr F}=\{F_0,F_1\}$ ($F_0,F_1\ge0$ and $F_0+F_1=\openone$).  Hypothesis~$H_0$ ($H_1$) is accepted iff $F_0$~($F_1$)~clicks. Two error probabilities are defined: $\alpha=\tr(F_1\rho)$, false-positive or type-I error; and $\beta=\tr(F_0\sigma)$, false-negative or type-II error. Generically, a decrease in type I error results in an increase in type-II error and vice versa.
%
%
Depending on the problem at hand, one may need to know, e.g.,~$\beta$ for a maximum allowed value of~$\alpha$; or one may instead be interested in the average error probability $p_{\rm err}=(\alpha+\beta)/2$, where for simplicity we assume equal priors for $H_0$ and $H_1$. This second possibility is known as  symmetric hypothesis testing  and leads to minimum error state discrimination~\cite{H1976}, where  $p_{\rm err}$ is minimized over all POVMs~$\mathscr F$.


When $H_0$, $H_1$ refer to  $\rho^{\otimes n}$, $\sigma^{\otimes n}$, i.e., to $n$ i.i.d. copies of $\rho$, $\sigma$, the error probabilities generically fall off exponentially as $n$ increases. 
Then, one is usually interested in the corresponding exponential rates. For symmetric hypothesis testing, the optimal rate is
given by the quantum Chernoff bound~\cite{ACMBMAV07, NS06}.
Similarly, the exponential rate of $\beta$ is  given by Stein's Lemma~\cite{HP91,ON00} if an upperbound is set on~$\alpha$,  
or by  Hoeffding's bound~\cite{H07,OH04,N06}  if instead  a lowerbound is set on the exponential rate of $\alpha$~\footnote{ The errors and rates for classical hypothesis testing can be recovered from those above by simply taking the matrices $\rho$ and  $\sigma$ to be diagonal with entries given by two probability distributions $P=\{P_k\}$, ${\overline P}=\{{\overline P}_k\}$, associated to~$H_0$ and~$H_1$ respectively
.
}. 
These asymptotic bounds have found many useful applications in quantum information theory, such as  providing an alternative proof of the classical capacity of a quantum channel~\cite{add1,WR12}, giving operational meaning to abstract quantities~\cite{CHMOSWW16}, quantum reading~\cite{Pirandola2011}, or in quantum illumination~\cite{Lloyd1463,lloyd08e}. 
%

Coming back to our original problem, we wish to assess the discrimination power of a device given by a specific POVM, ${\mathscr E}=\{E_1,\dots,E_m\}$. Let us assume that the positive operators $E_i$ (generically non-orthogonal) act on a finite $d$-dimensional Hilbert space, ${\mathscr H}_d$~\footnote{For simplicity and to ease up the notation we will assume POVM with a finite number of outcomes, however the results holds for any  POVM, including those with continuos outcome.}, of the quantum system $S$.
First, using free operations, we need to produce a valid POVM, ${\cal F} = \{F_0, F_1\}$, out of ${\mathscr E}$, to discriminate two states $\rho$ and $\sigma$. This can be achieved~\cite{OCMB11} by grouping (post-processing) the measurement outcomes,~$\{1,2,\dots,m\}$, in two disjoint sets $a$, $\bar a$,  and defining $E^{a}:=\sum_{k\in a} E_k$, $E^{\bar a}:=\sum_{k\in \bar a} E_k=\openone-E^a$. Then, ${\mathscr F}=\{E_{\mathscr{M}}^a,E_{\mathscr{M}}^{\bar a}\}$, where $E^{a}_{\mathscr{M}}=\mathscr{M}^{\dagger}(E^{a})$ (likewise for $E_{\mathscr{M}}^{\bar a}$), for a suitable trace preserving quantum operation    $\mathscr{M}$  (pre-processing). 
 The error probabilities thus read $\alpha=
\tr(E_{\mathscr{M}}^{\bar a} \rho)$ and~$\beta=
\tr(E_{\mathscr{M}}^{ a} \sigma)$.

In this single-shot scenario, we can now quantify the discrimination power of~$\mathscr E$ by the minimum average error probability. It~can be written as
\begin{equation}
p^*_{\rm err} = \min_a\min_{(\rho,\sigma)}\frac{1}{2} \big\{1 + \tr[E^{a}( \sigma -  \rho)]\big\},
\label{p_err}
\end{equation}
where the minimization is over all partitions $\{a,\bar a\}$ of the outcome set (over all post-processing operations) and over all state pairs $(\rho,\sigma)$, so $\mathscr{M}$ can be dropped in the minimization.
One can readily check~\cite{OCMB11} that the minimum single-shot error probability is given by the spread of $E^{a}$: $p^*_{\rm err} =1/2- \min_a (\lambda^{a}_{\mathrm{max}}-\lambda^{a}_{\mathrm{min}})/2$. This value is attained when~$\rho$ ($\sigma$) is the eigenstate of~$E^{a}$ corresponding to its maximum (minimum) eigenvalue.  


The single-shot scenario above is too restrictive since one can easily envision discrimination settings where the measurement $\mathscr E$ is performed a number~$n$ of times. In the most general setting, a system consisting of $n$ copies of $S$ is prepared (by, say, Alice) in one of the states of the pair $(\rho^n,\sigma^n)$, corresponding respectively to hypotheses~$H_0$ and~$H_1$. Here, $\rho^n, \sigma^n\in{\mathscr S}({\mathscr H}_d^{\otimes n})$, where ${\mathscr S}({\mathscr H})$ stands for the set of density matrices on a Hilbert space ${\mathscr H}$, can be fully general, not just of the form $\rho^{\otimes n}$, $\sigma^{\otimes n}$. The measurer's (say, Bob's) goal is to tell which hypothesis is true by  performing~$n$ measurements, all of them given by the POVM~$\mathscr E$. 
Free operations include again pre-processing of  $(\rho^n,\sigma^n)$ and post-processing of the classical data gather after each measurement. 
%
As in Eq.~(\ref{p_err}), when minimizing over state pairs, it is enough to choose the discriminating POVM as \mbox{${\mathscr F}=\{E^a,E^{\bar a}=\openone-E^a\}$}, where $E^a$ has now  the form
\begin{equation}
E^{a}= \sum_{{\bf k}^n \in a}E_{{\bf k}^n}:= \sum_{{\bf k}^n \in a} \bigotimes_{i=1}^n E_{k_i}.
\label{Ea n}
\end{equation}
Here ${\bf k}^r:=\{k_1,k_2,\dots,k_r\}$ denotes a 
sequence of outcomes of length $r$ (${\bf k}^0:=\emptyset$), so ${\bf k}^n$ is obtained after completing all measurements.  The two disjoint sets~$a$ and~$\bar a$ now contain all the sequences  assigned to the hypotheses $H_0$ and $H_1$ respectively.
\mbox{Type-I} and type-II error probabilities are $\alpha_{n} = \tr(E^{\bar a} \rho^{n})$ and $\beta_{n} = \tr(E^{ a} \sigma^{n})$, and the error probability for symmetric hypothesis testing can be written as~$p_{\rm err}=\min_{a}(\alpha_n+\beta_n)/2$~\footnote{The corresponding asymptotic error rate, Eq.~(\ref{zeta}), is independent of the prior probabilities $\eta_1$ and $\eta_2$, provided they do not vanish. Here we take $\eta_1=\eta_2=1/2$ for simplicity.}.

It is not hard to see that the errors fall off exponentially with $n$. It is then natural to quantify the discrimination power of $\mathscr E$ by the optimal asymptotic exponential rate of $p_{\rm err}$, defined as
\begin{equation}
\zeta =- \min_{(\rho^n,\sigma^n)} \;\lim_{n\rightarrow\infty} \frac{1}{n} \log{p_{\rm err}} .
\label{zeta}
\end{equation}

Although $p_{\rm err}$ can still be written as the spread of the optimal grouping, the number of groupings grows super-exponentially with $n$. Moreover, very little is known about the spectrum of operator sums such as those in Eq.~(\ref{Ea n}) and their eigenvectors (i.e.,~$\rho^n$ and~$\sigma^n$). A few examples that show the difficulties of dealing with finite $n$ are given below. We will thus evaluate Eq.~(\ref{zeta}) following an alternative route:   
We will prove that the exponential rate in Eq.~(\ref{zeta}) can be attained using i.i.d. states.
The main ingredient is to show that our problem is a particular case of classical channel discrimination. We will then use Ref.~\cite{H08}  to complete the proof.

\begin{figure}
\includegraphics[width=25em]{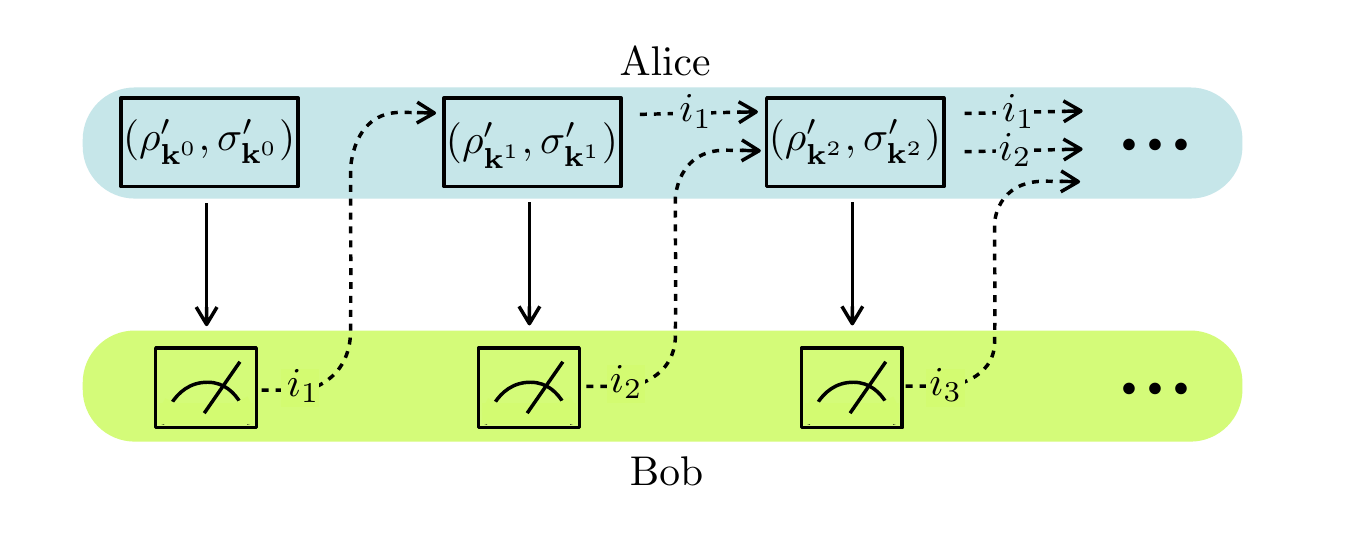}
\vspace{-2em}
\caption{Adaptive protocol. At each step (left to right), Alice sends to Bob (solid arrows) the state in Eq.~(\ref{rho'}), which she has  prepared using Bob's feedback (dashed arrows).}\label{Fig:adap1}
\end{figure}

To this end, let us momentarily broaden the scope of our original problem. First, we view hypothesis testing as a communication protocol where Alice (the state preparer) sends one of two possible messages, $H_0$,~$H_1$, to~Bob (the measurer) using suitable states in~${\mathscr S}({\mathscr H}_d^{\otimes n})$. Bob is allowed to perform~$n$ measurements with his detector to identify with minimum error which of the messages Alice sent. 
Second, in this communication context it is natural to allow classical feedback from Bob to Alice after each measurement. This enables an adaptive protocol (see Fig.~\ref{Fig:adap1}) in which
Alice sends one state at a time to Bob's detector and waits for him to provide feedback on the obtained outcome. Alice uses this information to prepare the succeeding state in a way that minimizes the identification error. 
Such protocols are widely used in quantum information theory~\cite{DFLS16, BSST99,PL16,GM2000,hayashi_comparison_2011}, particularly in quantum channel discrimination~\cite{HHLW09,Cooney2016}. 


It follows from the structure of 
$E_{{\bf k}^n}$ in Eq.~(\ref{Ea n}), that for {\em any} \mbox{$\rho^n\in{\mathscr S}({\mathscr H}_d^{\otimes n})$} (analogously for $\sigma^n$) there is an adaptive protocol that gives the same outcome probability distribution. Adaptive protocols are thus more general than those in which $\rho^n$ is entangled, so the optimal protocol can be chosen to be adaptive with no loss of generality. 

To prove the statement above, we define $\rho'_\emptyset:=\tr_{[n] \setminus 1}(\rho^n)$, where we denote by $[n]\!\!\setminus\!\! s$ the set $\{1,2,\dots,s-1,s+1,\dots n\}$, $s=1,2,\dots n$. Then,~$\rho^n$ and  $\rho'_\emptyset$ give the same probability distribution to the outcomes of Bob's first measurement: $P({k_1}|\rho^n):=\tr[(E_{k_1}\otimes\openone)\rho^n]=\tr(E_{k_1}\rho'_\emptyset)$. With Bob's feedback (the value of $k_1$), Alice can next prepare the second (unnormalized) state as $\rho'_{k_1}:=\tr_{[n]\setminus 2} [(E_{k_1}\otimes\openone)\rho^n]$. So, $\rho^n$ and $\rho'_{k_1}$ give the same outcome probabilities up to Bob's second  measurements: $P({{\bf k}^2}|\rho^n):=\tr [(E_{{\bf k}^2}\otimes\openone) \rho^n]=\tr(E_{k_2}\rho'_{k_1})$. 
Note that the probabilities of previous outcomes are implicit in the normalization of $\rho'_{k_1}$.
We readily see that if Alice's preparation at an arbitrary step $s$ is
\begin{equation}
\rho'_{{\bf k}^{s-1}}:=\tr_{[n]\setminus s}\left[\left( E_{{\bf k}^{s-1}}\otimes\openone\right) \rho^n\right],
\label{rho'}
\end{equation}
where we used the convention $E_{{\bf k}^0}=E_{\emptyset}:=\openone$, then $P({{\bf k}^{s}}|\rho^n)
=
\tr (E_{k_s}\rho'_{{\bf k}^{s-1}})$, $s=1,2,\dots,n$ (obviously, the analogous relation holds for $\sigma^n$, $\sigma'_{{\bf k}^{s-1}}$). This completes the proof.
 
We next show that the adaptive communication protocols introduced above, the optimal one in particular, can be cast as discrimination of two classical channels. 
To show this, we choose the classical input alphabet ${\cal X}$ to be any parametrization of ${\mathscr S }({\mathscr H}_d)\times {\mathscr S} ({\mathscr H}_d)$ (e.g., the set of all two Bloch vectors of $\rho$ and $\sigma$). 
The output alphabet  $\cal K$  is naturally given by the outcome labels of our fixed measurement (i.e., the POVM $\mathscr E$): ${\cal K}=\{1,2,\dots m\}$. 
We can then associate the null [alternate] hypothesis $H_0$ [$H_1$]  with the classical channel $W_x(k):= \tr (E_k \rho)$ [${\overline W}_x(k):= \tr (E_k \sigma)$], where  $(\rho,\sigma)=x\in{\mathcal X}$ and  $k\in{\cal K}$.
%
%
The optimal error rate is obtained by minimizing over~$x\in\cal X$.
%
%
So our single-shot problem can be seen as discrimination of the two classical channels $W$ and  $\overline W$ with one channel use.

This analogy holds also for our general, multiple-shot problem. The adaptive protocol defined by the states $\rho'_{{\bf k}^{s-1}},\, \sigma'_{{\bf k}^{s-1}}\in{\mathscr S}({\mathscr H}_d)$, $s=1,2,\dots, n$, 
translates into an adaptive channel discrimination strategy where~$n$ uses of $W$ or~$\overline W$ are allowed. 
%
We can now invoke the main result in~\cite{H08}. It states that for the problem of classical channel discrimination no adaptive strategy can outperform the best non-adaptive or fix strategy.  More precisely, it states that the optimal error rate can be attained  by the simple sequence where all the letters are equal,~$x_1=x_2=\dots=x_n$. We hence conclude that the optimal rates for our original problem can be achieved by i.i.d. state pairs, $(\rho^{\otimes n},\sigma^{\otimes n})$.
This holds for the Chernoff bound, Hoeffding's bound and for Stein's Lemma.

Computing the exponent rate $\zeta$ in Eq.~(\ref{zeta}) is now identical to computing the analogous rate for the classical hypothesis testing problem of discriminating between the probability distribution  $P_k=\tr(E_k \rho)$ and ${\overline P}_k =\tr(E_k \sigma)$ after $n$ samplings. It is given by the classical Chernoff Bound (CB)~\cite{Note1}:
\begin{equation}
\zeta_{\rm CB}=-\min_{(\rho,\sigma)}\;\min_{0\leq s\leq 1}\phi(s | P\|{\overline P}), 
\label{zeta =}
\end{equation}
where $\phi(s|P\|{\overline P}):=\log\sum_k P_k^s{\overline P}_k{}^{\!\!1-s}$. Eq.~(\ref{zeta =}) [along with Eq.~(\ref{other}) below] is the main result of this letter. A simple, very explicit, alternative proof of this result for two-dimensional two-element POVMs can also be found in~\cite{appendix}. Also in~\cite{appendix}, the reader will find upper and lower bounds to $\zeta_{\rm CB}$ (and to $\zeta_{\rm SL}$ and $\zeta_{\rm HB}$, defined below) that hold under mixing of POVMs.

It is plausible that Eq.~(\ref{zeta =}) is minimized by an orthogonal pair, as such pairs arguably give rise to the most distinguishable $P$ and $\overline P$ distributions. This would entail a simplification in the evaluation of $\zeta_{\rm CB}$.  
We have so far failed to prove the optimality of orthogonal pairs in full generality, though it is supported by extensive numerical analysis and  it holds for the examples in the next paragraph~\footnote{However, one can easily find counterexamples of the optimality of orthogonal pairs for the statistical overlap, given by $-\min_{(\rho,\sigma)}  \phi(\frac{1}{2},P |\kern-.1em| {\overline P})$. 
}.

To illustrate our results, in~\cite{appendix} we compute the discrimination power of the qubit covariant POVM, ${\mathscr E}=\{\openone+\mbox{\boldmath $n$}\cdot \mbox{\boldmath$\sigma$}\}_{\mbox{\boldmath\scriptsize$n$}\in{\mathbb S}^2}$, where~$\mbox{\boldmath$\sigma$}$ is the vector of Pauli matrices and ${\mathbb S}^2$ is the unit 2-sphere. The result is $\zeta_{\rm CB}=-\log(\pi/4)$ which can be compared to $\zeta_{\rm CB}=-(1/2)\log(1-r^2)$, corresponding to a noisy Stern-Gerlach of purity $r$. We see that  ${\mathscr E}$ has the same discrimination power that a Stern-Gerlach with purity $r \approx 0.62$.

So far,  special emphasis has been placed on the asymptotics of the problem at hand. 
It is illustrative to examine with a few examples the difficulties arising for finite $n$, where some of the asymptotic results do not hold. Let us focus on two-element POVMs, ${\mathscr E}=\{E_1,E_2=\openone-E_1\}$. In this case, $E_1$ and $E_2$ commute and can be diagonalized simultaneously. In the multiple-shot  scenario, the groupings $\{E^{a},E^{\bar a}\}$ will also be diagonal in the very same local basis that diagonalize $E_1$ and $E_2$ and thus each state of the optimal pair, $(\rho^n,\sigma^n)$, in necessarily a product state of elements of that basis.  In this case, however, one can show that i.i.d. states are not necessarily optimal. In~\cite{appendix}  we give a concrete example for~$n=3$ where the optimal states are~$\rho^3=|001\rangle\langle001|$ and  $\sigma^3=|110\rangle\langle110|$, rather than~$|000\rangle\langle000|$ and $|111\rangle\langle111|$. Furthermore, we also show that there exists an adaptive protocol with yet a smaller error rate, 
thus outperforming the optimal non-adaptive protocol for $n=3$.

Though in this letter we have focused on the problem dual to symmetric hypothesis testing, which led us to Eq.~(\ref{zeta =}), the very same arguments concerning the optimality of i.i.d. state pairs apply to the dual Stein's lemma and Hoeffding's bound,
%
whose asymptotic rates are defined as \mbox{$\zeta_{{\rm SL}/{\rm HB}}=-\min_{(\rho^n,\sigma^n)}\lim_{n\to\infty}(1/n)\log \beta_n$}, where the minimization is subject to $\alpha_n\le\epsilon$  and $\alpha_n\le{\rm e}^{-nr}$  respectively. It follows from our analysis that they can be computed simply as
\begin{eqnarray}
\zeta_{\rm SL}\!&=&\max_{(\rho,\sigma\!)} D(P\|{\overline P}),\nonumber\\
\zeta_{\rm HB}\!&=&\max_{(\rho,\sigma\!)}\,\sup_{0\le s\le 1}\!\!{-s r - \phi( s|P\|{\overline P})\over 1-s},
\label{other}
\end{eqnarray}
where $D(P\|{\overline P})=\sum_k P_k\log(P_k/{\overline P}_k)$ is the relative entropy.
In summary, we have introduced a class of problems dual to quantum hypothesis testing where the measurement device is a given. We have derived simple expressions for the asymptotic (exponential) error rates, which quantify the discrimination power of the measurement device when it can be used multiple times. We have briefly discussed two paradigmatic examples for qubits, covariant POVMs and noisy Stern-Gerlach, and addressed the non-asymptotic regime with some examples. As final remarks, we point out that these dual problems complement our understanding of quantum hypothesis testing and state discrimination and we believe they will find many applications in quantum information theory. Open problems include a deeper understanding of the structure of the optimal pairs and extending the analysis to infinite dimensional systems such as light-fields, where constraints on the mean energy are necessary.

%

\paragraph*{Acknowledgments}
CH, EB and JC acknowledge support by the Spanish MINECO, project FIS2013-40627-P and CH by FPI Grant No. BES-2014-068888, as well as by the Generalitat de Catalunya, CIRIT project no. 2014 SGR 966.
MH is partially supported by a MEXT Grant-in-Aid for Scientific Research (A) No. 23246071 and the Okawa Reserach Grant. Centre for Quantum Technologies is a Research Centre of Excellence funded by the Ministry of Education and the National Research Foundation of Singapore.

\bibliographystyle{apsrev4-1}
\bibliography{Bib}

\clearpage
\setcounter{equation}{0}
\begin{widetext}
\setcounter{page}{1}
\appendix

\section{Appendix: Additional Remarks}

\section{Some properties of the error exponents}

In this section we derive some properties of the optimal error exponents. 
To this end, we investigate the function
\begin{align}
C_{\mathscr E}&=\min_{(\rho,\sigma)}\min_{0\leq s\leq 1} \sum_{i=1}^{m} \left[\tr(\rho E_i)\right]^s \left[\tr(\sigma E_i)\right]^{1-s},
\label{C_e}
\end{align}
for a given $m$-element POVM, ${\mathscr E}=\{E_i\}_{i=1}^m$. We will denote the corresponding error exponents in the three settings discussed  in the letter by $\zeta^{\mathscr E}_{\rm CB}$, $\zeta^{\mathscr E}_{\rm SL}$ and $\zeta^{\mathscr E}_{\rm HB}$.  
Let us investigate the behavior of the discrimination power under mixing of POVMs.
\begin{lemma}
Let ${\mathscr E}$ be a POVM with $m$ elements $E_i$ and ${\mathscr G}$ be a POVM with $n$ elements $G_i$. Define a mixed POVM with $m+n$ elements, $\widehat{\mathscr E}=\{\widehat E_i\}_{i=1}^{m+n}$, through
\begin{equation}
\widehat{ E}_i=\left\{
\begin{array}{rcl}p E_i&\mbox{if}& 1\le i\le m,\\[.5em]
(1-p)G_{i-m} &\mbox{if}& m< i\le m+n .
\end{array}
\right.
\end{equation}
Then, we can upper bound $\zeta_{\rm CB}^{\widehat {\mathscr E}}$ by
\begin{equation}
\zeta_{\rm CB}^{\widehat {\mathscr E}} \leq p \zeta_{\rm CB}^{\mathscr E} + (1-p) \zeta_{\rm CB}^{\mathscr G}
\end{equation}
and lower bound it by 
\begin{equation}
\zeta_{\rm CB}^{\widehat {\mathscr E}} \geq - \log \min \Big\{ p C_{\mathscr E} + (1-p) , p + (1-p) C_{\mathscr G} \Big\}.
\label{lw CB}
\end{equation}
%
Furthermore we can state the following relations for Stein's bound and Hoeffding's bound
\begin{align}
\zeta_{\rm SL}^{\widehat {\mathscr E}} \leq p\, \zeta_{\rm SL}^{\mathscr E} + (1-p) \zeta_{\rm SL}^{\mathscr G}, \\[.5em]
\zeta_{\rm SL}^{\widehat {\mathscr E}} \geq \max \left\{ p\, \zeta_{\rm SL}^{\mathscr E} ,  (1-p) \zeta_{\rm SL}^{\mathscr G} \right\}, \\[.5em]
\zeta_{\rm HB}^{\widehat {\mathscr E}} \leq p\, \zeta_{\rm HB}^{\mathscr E} + (1-p) \zeta_{\rm HB}^{\mathscr G}.
\end{align}
\end{lemma}

\begin{proof}
Define ${\mathscr E}$, ${\mathscr G}$ and $\widehat {\mathscr E}$ as above. Let us first give a lower bound for $C_{\widehat{\mathscr E}}$.
\begin{align}
C_{\widehat {\mathscr E}} &= \min_{(\rho,\sigma)}\min_{0\leq s\leq 1} \sum_{i=1}^{m+n} \left[\tr(\rho \widehat E_i)\right]^s \left[ \tr(\sigma \widehat E_i)\right]^{1-s} \nonumber\\[.5em]
&\geq \min_{(\rho,\sigma)}\min_{0\leq s\leq 1} \sum_{i=1}^{m} \left[\tr(\rho \widehat E_i)\right]^s \left[ \tr(\sigma \widehat E_i)\right]^{1-s} + \min_{(\rho,\sigma)}\min_{0\leq s\leq 1} \sum_{i=m+1}^{m+n} \left[\tr(\rho \widehat E_i)\right]^s \left[\tr(\sigma \widehat E_i)\right]^{1-s}  \nonumber\\[.5em]
&= \min_{(\rho,\sigma)}\min_{0\leq s\leq 1} \sum_{i=1}^{m} p\left[ \tr(\rho E_i)\right]^s \left[\tr(\sigma  E_i)\right]^{1-s} + \min_{(\rho,\sigma)}\min_{0\leq s\leq 1} \sum_{i=1}^{n} (1-p) \left[\tr(\rho G_i)\right]^s\left[ \tr(\sigma G_i)\right]^{1-s}  \nonumber\\[.5em]
&= p C_{\mathscr E} + (1-p) C_{\mathscr G}.
\end{align}
We continue by giving an upper bound to $C_{\widehat{\mathscr E}}$. 
Let $s^*$, $\rho^*$ and $\sigma^*$ be respectively the value of $s$ and the states~$\rho$ and $\sigma$ that attain the minimum value on the right hand side of Eq.~(\ref{C_e}) for the POVM $\mathscr E$. Then, 
\begin{align}
C_{\widehat {\mathscr E}} 
= \min_{(\rho,\sigma)}\min_{0\leq s\leq 1} \sum_{i=1}^{m+n} \left[\tr(\rho \widehat E_i)\right]^s \left[\tr(\sigma \widehat E_i)\right]^{1-s}
\leq p\, C_{\mathscr E} + \sum_{i=m+1}^{m+n}\left[ \tr(\rho^* \widehat E_i)\right]^{s^*} \left[\tr(\sigma^* \widehat E_i)\right]^{1-s^*}.
\end{align}
We prove one of the lower bounds by bounding the second term as 
\begin{align}
\sum_{i=m+1}^{m+n}\left[ \tr(\rho^* \widehat E_i)\right]^{s^*} \left[\tr(\sigma^* \widehat E_i)\right]^{1-s^*}
\leq&  (1-p) \left[ \sum_{i=1}^{n} \tr(\rho^* G_i)\right]^{s^*} \left[ \sum_{i=1}^{n} \tr(\sigma^* G_i) \right]^{1-s^*} 
= 1-p,
\end{align}
where the inequality follows from the definition of $\widehat {\mathscr E}$ and H\"{o}lder's inequality. An analogous bound follows by choosing  $s^*$, $\rho^*$ and $\sigma^*$ to be  the value of $s$ and the states $\rho$ and $\sigma$ that attain the minimum value on the right hand side of Eq.~(\ref{C_e}) for the POVM $\mathscr G$.  Hence, 
$C_{\widehat {\mathscr E}} \le \min\left\{p\, C_{\mathscr E}+(1-p),p+(1-p)C_{\mathscr G}\right\}$, and Eq.~(\ref{lw CB}) follows.
The inequalities for Steins and Hoeffdings bounds are proven similarly, using additionally that the relative entropy is lower bounded by zero and  that the logarithm is concave. 
\end{proof}

The above shows that by mixing a pair of POVMs one can never increase the discrimination power of the best POVM of the pair. Furthermore, it is easy to see that applying additional CPTP maps before performing the measurement does not increase the discrimination power either. 
This follows directly from the fact that the image of a CPTP map is always at most the input state space itself. 

\section{Example for finite number of measurements}

Let us consider the simple example where the POVM is
\begin{equation} \label{POVMexam}
{\mathscr E} = \left\{ E_1 = \left(
\begin{array}{cc}
0.4&0\\
0&0.2
\end{array}\right), E_2 = 1-E_1 = \left(
\begin{array}{cc}
0.6&0\\
0&0.8
\end{array}\right)\right\},
\end{equation}
and $n=3$ (the measurement defined by $\mathscr E$ is performed 3 times). It follows from the diagonal form of $E_1$ and $E_2$ that the optimal input states are tensor products of pure states, also diagonal in the given basis. From the symmetry of the problem it should be clear that there are only two possible ways to achieve the optimal error rate: 
(i)~use the pair $(\rho_0^{\otimes 3},\rho_1^{\otimes 3})$, or (ii)~use $(\rho_0^{\otimes 2}\otimes \rho_1,\rho_1^{\otimes 2}\otimes \rho_0 )$, where 
\begin{equation}\label{rho 0 rho 1}
\rho_0 =
\begin{pmatrix}
1&0\\
0&0
\end{pmatrix},
\qquad 
\rho_1 = 
\begin{pmatrix}
0&0\\
0&1
\end{pmatrix}.
\end{equation}
We next compute the error probability in both cases. For (i) it can be checked that the optimization over groupings gives
\begin{equation} 
F^{\rm (i)}_1 = E_2 \otimes E_2 \otimes E_2,\quad  F^{\rm (i)}_0 =\openone-F^{\rm (i)}_1,
\end{equation}
so $F^{\rm (i)}_0$ is the sum of the remaining seven tensor products. 
One can easily check that the error probability is
\begin{equation}
p_{\rm err}^{\rm(i)} = 0.352. 
\end{equation}
For (ii), the optimization over groupings  gives now
\begin{align} 
F_0^{\rm(ii)} = E_1 \otimes E_1 \otimes E_1 + E_1 \otimes E_1 \otimes E_2
+ E_1 \otimes E_2 \otimes E_2  + E_2 \otimes E_1 \otimes E_2 ,
\end{align}
and $F_1^{\rm(ii)}  = 1 - F_0^{\rm(ii)} $, being the sum of the remaining four products.
This gives
\begin{equation}\label{err1eq}
p_{\rm err}^{\rm(ii)} = 0.344. 
\end{equation}
We see that (ii) is optimal. The optimal state pair in this example for finite $n$ is not of the form $(\rho^{\otimes n},\sigma^{\otimes n})$, in contrast with what we found in the asymptotic limit of large $n$. 

Furthermore, we next show that the error rate given in Eq.~(\ref{err1eq}) can be lowered by an adaptive protocol as follows. We choose the first two input pair to be, as in the previous examples, $(\rho_0^{\otimes 2},\rho_1^{\otimes 2})$.
 If the first and second measurement return $1$ (i.e., if  ${\bf k}^2=\{1,1\}$),  the preparation of the third state pair is $(\rho_0,\rho_1)$, and we swap preparations otherwise [i.e., the third pair is $(\rho_1,\rho_0)$, as in~(ii)].
One can check that the optimal grouping is 
\begin{align} 
F^{\rm ad}_0 = E_1 \otimes E_1 \otimes E_1 + E_1 \otimes E_1 \otimes E_2 
+ E_2 \otimes E_2 \otimes E_1 + E_1 \otimes E_2 \otimes E_2  + E_2 \otimes E_1 \otimes E_2 ,
\end{align}
and $F^{\rm ad}_1 = 1 - F^{\rm ad}_0$.
This results in an error probability of
\begin{equation}\label{err1eq2}
p_{\rm err}^{\rm ad} = 0.336. 
\end{equation}

\begin{figure}[tp]
\centering
\includegraphics[scale=0.7]{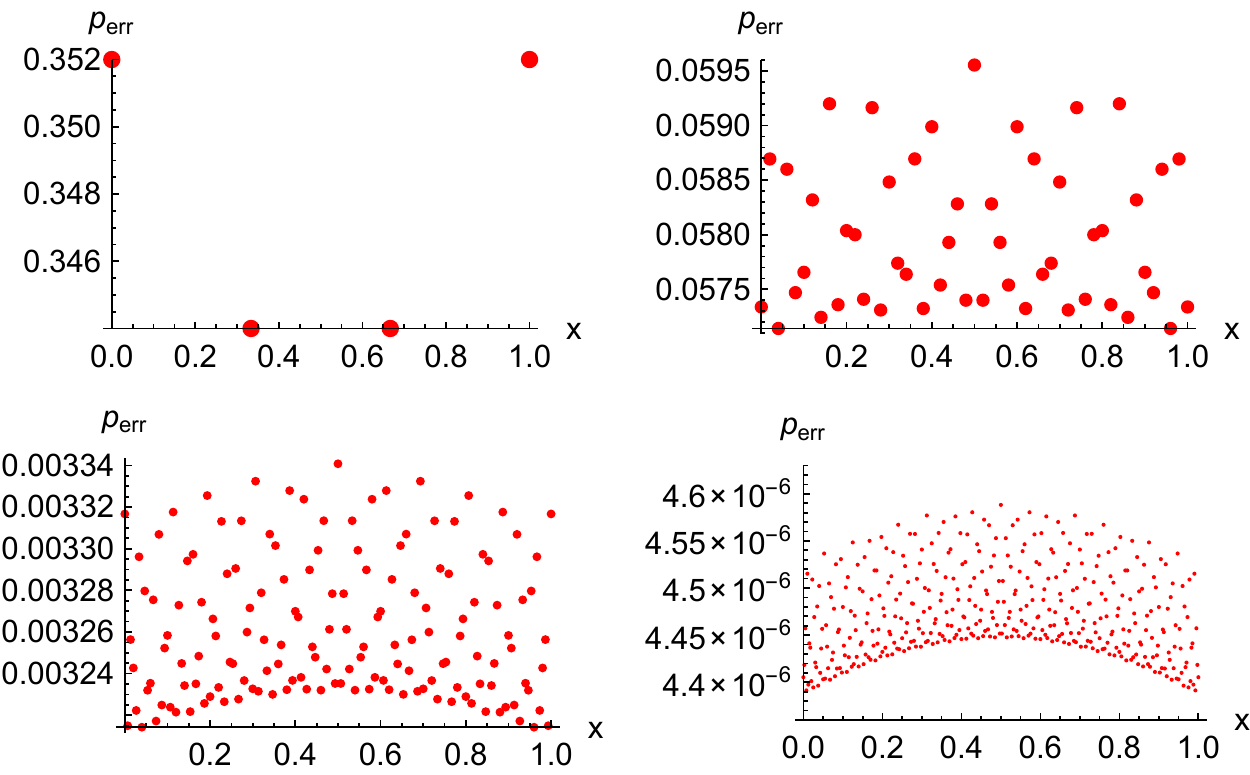}
\caption{Error probability vs. $x$ for state pairs of the form $(\rho_0^{\otimes xn}\otimes\rho_1^{\otimes(1-x)n},\rho_1^{\otimes xn}\otimes\rho_0^{\otimes(1-x)n})$. The values of $n$ are~$3$, $50$, $150$, and $400$ (from top left to lower right), with the POVM given as in Eq.~(\ref{POVMexam}).  \label{fig:POVMnm} }
\end{figure}

Equation~(\ref{err1eq}) provides an explicit example that non i.i.d. states can outperform i.i.d. ones for finite number of measurements.
Nevertheless, we proved in the main text that asymptotically the choice $(\rho_0^{\otimes n},\rho_1^{\otimes n} )$ is optimal. Figure~\ref{fig:POVMnm} illustrates that this is indeed so by showing plots of the error probability arbitrary state pairs. Since  permutations among subsystems do not affect the error probability, the state pairs can be taken to be without loss of generality of the following form:
\begin{equation}\label{rho x}
(\rho^n,\sigma^n)=\left(\rho_0^{\otimes xn}\otimes\rho_1^{\otimes(1-x)n},\rho_1^{\otimes xn}\otimes\rho_0^{\otimes(1-x)n}\right).
\end{equation}
As $n$ increases, we see that the minimum error probability becomes a convex function of $x$.
This is formally proven below (in section \ref{sec:explic}), where we explicitly compute the optimal rates for 2-outcome qubit POVMS.

\section{Covariant measurements}

In this section we prove that orthogonal states are optimal for the qubit covariant POVM,
\begin{equation}
{\mathscr E}=\left\{E_{\scriptsize\vec n}:= \openone+\vec n\cdot\vec\sigma\right\}_{{\scriptsize\vec n}\in{\mathbb S}^2},
\end{equation}
and we compute its discrimination power according to the dual of the Chernoff bound.
For the covariant measurement $\mathscr E$, we have
\begin{equation}
C_{\mathscr E}=\min_{0\le s\le 1}C_s,
\end{equation}
 where
\begin{equation}
C_s=\min_{(\rho,\sigma)}\int_{{\mathbb S}^2} dn\left[\tr\left(\rho E_{\scriptsize\vec n}\right)\right]^s\left[\tr\left(\sigma E_{\scriptsize\vec n}\right)\right]^{1-s}=\min_{\mbox{\scriptsize$(\vec m_1$},\mbox{\scriptsize$\vec m_2)$}} C,
\end{equation}
and
\begin{equation}
C:=\int   dn  \left( {1 +\cos\widehat{\vec n\,\vec m}_1}  \right)^{  1-s}   \left( {1 +\cos\widehat{\vec n\,\vec m}_2} \right)^{  s} .
\end{equation}
Here $\vec m_1$ and $\vec m_2$ are the Bloch vectors of $\sigma$ and $\rho$ respectively.
Chosing with no loss of generality
\begin{equation}
\vec m_1=\hat z,\quad 
\vec m_2= \hat z \sin\alpha + \hat x \cos\alpha,
\end{equation}
where $\hat z$ and $\hat x$ are the unit vectors pointing in the $z$ and $x$ direction
($\vec m_1$ and $\vec m_2$ on the $xz$-plane), we have
\begin{equation}
C=\int_0^\pi \sin\theta \int_0^{2\pi}{d\phi\over4\pi} \left(1+\cos\theta \right)^{1-s}
\left(1+\cos\theta\cos\alpha+\cos\phi\sin\theta\sin\alpha \right)^{s}.
\end{equation}
where $\theta$ and $\phi$ are the polar and azimutal angles of the unit vector $\vec n$.
Using the relation
\begin{equation}
a^s={\sin(s\pi)\over\pi}\int_0^\infty
dx{a x^{s-1}\over a+x} 
\label{relation}
\end{equation}
in the second factor and making the usual change of variables, $\theta\to u:=\cos\theta$,
one has
\begin{equation}
C=\int_{-1}^{1} du \left(1+u\right)^{1-s}{\sin(s\pi)\over4\pi^2}\int_0^\infty dx
 \int_0^{2\pi}{d\phi\over4\pi}
{(1+u\cos\alpha+\sqrt{1-u^2}\cos\phi\sin\alpha)\,x^{s-1}\over
1+x+u\cos\alpha+\sqrt{1-u^2}\cos\phi\sin\alpha} .
\end{equation}
The $\phi$-integration can be carried out using the residue theorem, with the result
\begin{equation}
C=\int_{-1}^{1}    du \left(1+u\right)^{1-s}{\sin(s\pi)\over2\pi}\int_0^\infty dx\, x^{s-1}
\left[
1 - {x\over\sqrt{(1 + x + u\cos\alpha)^2 - (1 - u^2)\sin^2\alpha}}
\right].
\label{after x}
\end{equation}
Minimization over $\vec m_1$ and $\vec m_2$ is equivalent to minimization over $0\le\alpha\le \pi$. So,
let us take the derivative with respect to $\alpha$. One has
\begin{eqnarray}
{dC\over d\alpha}  &=&  -\sin\alpha{\sin(s\pi)\over2\pi}\int_0^\infty dx\, x^s\int_{-1}^{1}   du \left(1+u\right)^{1-s}
{(1+x)u+\cos\alpha\over\left[(1 + x + u\cos\alpha)^2 - (1 - u^2)\sin^2\alpha\right]^{3/2}}\nonumber\\[1em]
&:=&
  -\sin\alpha{\sin(s\pi)\over2\pi}\int_0^\infty dx\, x^s f(\alpha,x),
\label{def f}
\\ \nonumber
\end{eqnarray}
which shows that $\alpha=0,\,\pi$, are extreme points of $C$. To prove that these are the only extremes,
we next show that $f(\alpha,x)>0$ if $0<\alpha< \pi$. We first notice that the integrant in the first line of Eq.~(\ref{def f}) is
$$
-{1\over \sin^2\alpha}\; {d\over d u} {1+x+u \cos\alpha\over \sqrt{(1 + x + u\cos\alpha)^2 - (1 - u^2)\sin^2\alpha}}.
$$
Thus, integrating by parts one has
\begin{equation}
f(\alpha,x)={1 - s\over\sin^2 \alpha} \int_{-1}^1  \, {dx\,(1+u)^{-s}(1+x+u\cos\alpha)\over \sqrt{(1 + x + u\cos\alpha)^2 - (1 - u^2)\sin^2\alpha}}
-{2^{1-s}\over\sin^2\alpha}.
\end{equation}
We further note that if $0<\alpha<\pi$,
\begin{equation}
{1+x+u\cos\alpha\over \sqrt{(1 + x + u\cos\alpha)^2 - (1 - u^2)\sin^2\alpha}}>1.
\end{equation}
Hence
\begin{equation}
f(\alpha,x)>{1 - s\over\sin^2 \alpha}\int_{-1}^1dx\, (1+u)^{-s}-{2^{1-s}\over\sin^2\alpha}=0.
\end{equation}
Therefore, $dC/d\alpha\le 0$, and it only vanishes at $\alpha=0,\pi$. It follows that $C$ has a maximum at $\alpha=0$ and a minimum at $\alpha=\pi$.

Substituting $\alpha=0$ in Eq.~(\ref{after x}) (or alternatively, in the definition of $C$), we can write
\begin{eqnarray}
C_s&=&\int_{-1}^{1} du \left(1+u\right)^{1-s}{\sin(s\pi)\over2\pi}\int_0^\infty dx\, x^{s-1}
{1-u\over1+x-u}
={1\over2}\int_{-1}^{1} du \left(1+u\right)^{1-s}(1-u)^s\nonumber\\
&=&2\int_0^1dt\, t^{1-s}(1-t)^s=2B(2-s,1+s),
\end{eqnarray}
where we again have used Eq.~(\ref{relation}) (backwards), we have changed variables as $u\to t=(1+u)/2$, and used the definition of the Euler Beta function $B(a,b)$. Additionally, we know that
\begin{equation}
B(2-s,1+s)={\Gamma(2-s)\Gamma(1+s)\over \Gamma(3)}
={s(1-s)\over2}\Gamma(1-s)\Gamma(s)
={s(1-s)\pi\over2\sin(s\pi)} .
\end{equation}
We finally have that the Chernoff bound is
$
C_{\mathscr E}={\pi/4} 
$,
as $C_s$ has its minimum at $s=1/2$. It follows that the error exponent corresponding to the dual of the Chernoff bound is
\begin{equation}
\zeta_{\rm CB}=-\log(\pi/4).
\end{equation}

\section{Noisy Stern-Gerlach}

The noisy Stern-Gerlach measurement of purity $r$ is defined by
\begin{equation}
{\mathscr E}=\left\{{\openone+ r \sigma_z\over2},{\openone-r \sigma_z\over2}\right\}.
\end{equation}
Then, if $\vec m_1$, $\vec m_2$ are the Bloch vectors of $\sigma$ and $\rho$ respectively, one has
\begin{equation}
C={1\over2}\left(1+r \cos\theta_1\right)^{1-s}\left(1+r \cos\theta_2\right)^s+{1\over2}\left(1-r \cos\theta_1\right)^{1-s}\left(1-r\cos\theta_2\right)^s.
\end{equation}
One can easily check that $C$ attain its minima over the pair $(\theta_1,\theta_2)$ at $(0,\pi)$ and $(\pi,0)$, thus
\begin{equation}
C_s=\min_{\mbox{\scriptsize$(\vec m_1$},\mbox{\scriptsize$\vec m_2)$}}C=
{1\over2}\left(1\pm r \right)^{1-s}\left(1\mp r \right)^s+{1\over2}\left(1\mp r\right)^{1-s}\left(1\pm r\right)^s.
\end{equation}
The minimum over $s$ is at $s=1/2$, so we obtain
\begin{equation}
C_{\mathscr E}=\sqrt{1-r^2},\qquad \zeta_{\rm CB}=-{1\over2}\log(1-r^2).
\end{equation}
Note that $C_{\mathscr E}$ ($\zeta_{\rm CB}$) vanish (goes to $\infty$) for noiseless Stern-Gerlach apparatus, i.e.,  for $r=1$,  as $E_1$ and $E_{2}$, become orthogonal projectors and thus can be used for perfect discrimination.

\section{Discrimination with commuting qubit POVMs}\label{sec:explic}
Here, we consider a qubit two-elements POVM ${\mathscr E}=\{E_1,E_2\}$, such as that used in the numerical examples above,~Eq.~(\ref{POVMexam}). 
A simplification arises from the fact that~$E_1$ and~$E_2$ necessarily commute and can be diagonalized simultaneously, so with no loss of generality we can write
\begin{equation}
E_1 = \left(
\begin{array}{cc}
p&0\\
0&q
\end{array}\right) ;   \qquad      E_2 = \left(
\begin{array}{cc}
1-p&0\\
0&1-q
\end{array}\right);\qquad  p\geq q.
\end{equation}
as in Eq.~(\ref{POVMexam}).
We will prove that the intuition gathered from Fig.~\ref{fig:POVMnm} is correct and~i.i.d. states are optimal in the asymptotic limit by showing that the error exponent becomes a concave function of $x$ [the fraction of $\rho_0$ tensor powers in $\rho^n$, Eq.~(\ref{rho x})] with minima at $x=0$ and $x=1$. 
This is a particular case of our key result in the main text. Here, we provide a direct and intuitive proof in the special case under consideration. The expression for the asymptotic rate that we will derive below does not involve any optimization. The case at hand thus provides another example where the calculation of the dual Chernoff bound is straightforward. 

Our starting point is the analogous of Eq.~(1) in the main text for the multiple-shot scenario,
\begin{equation}
p^*_{\rm err} = \min_a\min_{(\rho^n,\sigma^n)}\frac{1}{2} \big\{1 + \tr[E^{a}( \sigma^n -  \rho^n)]\big\},
\end{equation}
where $E^a$ are given in Eq.~(2). As explained in the main text, $p^*_{\rm err}$ is given by the spread of $E^{a}$: 
\begin{equation}
p^*_{\rm err} ={1\over2}-{1\over2} \min_a (\lambda^{a}_{\mathrm{max}}-\lambda^{a}_{\mathrm{min}}).
\end{equation}
 This value is attained when~$\rho^n$ ($\sigma^n$) is the eigenstate of~$E^{a}$ corresponding to its maximum (minimum) eigenvalue $\lambda^{a}_{\mathrm{max}}$ ($\lambda^{a}_{\mathrm{min}}$).  
%
%
Because the two elements of the POVM commute, the states $\rho^n$ and $\sigma^n$ are of the form in Eq.~(\ref{rho x}). 
Then, if $m = n x$,
\begin{align}
\kern-.4em p^*_{\rm err}\! 
= \!\min_{0	\leq m\leq n}\!\min_a \frac{1}{2} \!
\left\{
\!1 \!- \!\!\sum_{i,j \in a} \!\!\binom{m}{i} \binom{n\!-\!m}{j}\!\Big[p^i(1\!-\!p)^{m-i}q^j(1\!-\!q)^{n-m-j} \!\!-\! q^i(1\!-\!q)^{m-i}p^j(1\!-\!p)^{n-m-j}\Big]\!\right\}\!.
\end{align}
Using large deviation (logarithmically tight) bounds we can approximate $p^*_{\rm err}$ in the asymptotic regime of large $n$ as 
\begin{align}
p_{\rm err} &\approx e^{-n\left[x D(s \|p) + (1-x) D(t \|q)\right]},
\end{align}
such that
\begin{equation}
x D(s \|p) + (1-x) D(t \|q) = x D(s \|q) + (1-x) D(t \|p) \label{equal},
\end{equation}
where 
\begin{equation}
s=s(x)=\frac{i}{xn}; \qquad     t=t(s,x)=\frac{j}{(1-x)n},
\end{equation}
and $D(s\| p)$ is the relative entropy, defined as
\begin{equation}
D(s\| p)=s\log{s\over p}+(1-s)\log{1-s\over1-p} .
\end{equation}
To ease the notation, we further define ${\mathscr D}:=x D(s \|p) + (1-x) D(t \|q)$. 
From Eq.~(\ref{equal}) we get
\begin{align}\label{gamma}
x  s - (1-x)  t = (2  x -1) \gamma,
\end{align}
where
\begin{align}
\gamma = \gamma(p,q) = {\displaystyle\log{\frac{1-q}{1-p}}\over \displaystyle \log{\frac{1-q}{1-p}}+\log{\frac{p}{q}}}.
\end{align}

Next, we have to minimize $ {\mathscr D}$ over~$s$, we thus impose the condition  ${\partial  {\mathscr D}}/{\partial s}=0$,  which implies
\begin{equation}\label{cond 1}
\frac{\partial D(t \|q)}{\partial t} = - \frac{\partial D(s \|p)}{\partial s}.
\end{equation}
In deriving this expression we have used that
\begin{equation}
{\partial t\over\partial s}={x\over 1-x},
\end{equation}
which follows from Eq.~(\ref{gamma}).
From the definition of $ D(t \|q)$, it also follows that
\begin{align}
\frac{\partial D(t \|q)}{\partial t} = \log{\frac{t}{q}\frac{1-q}{1-t}} .
\end{align}
Substituting this into Eq.~(\ref{cond 1}) we readily see that
\begin{align}
p  q   (1-s)  (1-t) = s  t (1-p)  (1-q) . \label{stx}
\end{align}
We next take the derivative of Eq.~(\ref{stx}) and Eq.~(\ref{gamma}) with respect to $x$ and obtain respectively the relations
\begin{align}
\frac{\partial s}{\partial x} &=- \frac{s+(1-s)r}{[s+(1-s)r]
{\displaystyle{ x\over\displaystyle 1-x}} + [t + (1-t)r]}  \frac{\partial t}{\partial x}:=Z \frac{\partial t}{\partial x},\label{Z} \\
{\partial t\over\partial x}&={s+t-2\gamma\over 1-x}={s-\gamma\over(1-x)^2},
\end{align}
where
\begin{equation} 
r=\frac{p  q}{(1-p)  (1-q)}.
\end{equation}
%

With this,  the calculation of $d  {\mathscr D}/d x$ and $d^2  {\mathscr D}/d x^2$ is straightforward. 
\begin{equation}
\frac{d  {\mathscr D}}{d x} = D(s \|p) - D(t \|q) + (1-x)\frac{\partial D(t \|q)}{\partial t} \frac{\partial t}{\partial x}. 
\end{equation}
One can easily check that $\left.d  {\mathscr D}/dx\right|_{x=1/2}= 0$. Hence, $ {\mathscr D}$ has an extreme value  at $x=1/2$.
Furthermore, 
\begin{align}
\frac{d^2  {\mathscr D}}{d x^2} &= \frac{1-x}{t  (1-t)} \left(\frac{\partial t}{\partial x}\right)^2 \left(1+\frac{x}{1-x}Z\right) \geq 0,
\end{align}
where $Z$ is defined in Eq.~(\ref{Z}).
%
Therefore, $ {\mathscr D}$ is a convex function of $x$ with minimum at $x=1/2$. It~follows that the optimal input state for the original problem is always given by either $x=0$ or $x=1$, which lead essentially to the same protocol because of the symmetry of the problem.

The (optimal) error exponent is thus given by 
\begin{align}
 \lim_{n\rightarrow\infty} - \frac{1}{n} \log{p^*_{\rm err}} = D(\gamma\|p),
\end{align}
which can be seen to be equal to $\zeta_{\rm CB}=-\min_s\phi(s|P\|{\overline P})$ for the binary distributions $P=\{p,1-p\}$, ${\overline P}=\{q,1-q\}$ (see for instance ~\cite{CMMAB08}).

\clearpage
\end{widetext}

\end{document}